\documentclass[ceqn, prd, showpacs, showkeys, prd, onecolumn, nofootinbib, amsmath, amsfonts, amssymb]{revtex4-2}

\usepackage{amssymb,amsmath,mathtools,xcolor,graphicx,xspace,colortbl,ragged2e,rotating} %
\usepackage{textcomp}
\usepackage{amsfonts}  
\usepackage{amsmath}  
\usepackage{amssymb}  
\usepackage{boxedminipage}  
\usepackage{graphics}  
\usepackage{graphicx}  
\usepackage{ragged2e}  
\usepackage{tabulary}  
\usepackage{wrapfig}  
\usepackage{xcolor}  
\usepackage{varioref}  
\usepackage{natbib}
\usepackage{revsymb}
\graphicspath{{FRACTIONAL8_1_graphics/}{FRACTIONAL8_1_tcache/}{FRACTIONAL8_1_gcache/}}
\DeclareGraphicsExtensions{.pdf,.eps,.ps,.png,.jpg,.jpeg}

\begin{document}
\title{FRACTIONAL-DIMENSION GRAVITY AND THE MILKY WAY GALAXY}
\author{Gabriele U. Varieschi}
\email[E-mail me at: ]{Gabriele.Varieschi@lmu.edu}
\homepage[Visit: ]{https://gvarieschi.lmu.build}
\affiliation{Department of Physics, Loyola Marymount University, Los Angeles, CA 90045, USA}
\date{\today}

\begin{abstract}
In this work, we focus our analysis of Fractional-Dimension Gravity (FDG) on our home galaxy, the Milky Way (MW), by using the latest Gaia DR3 data as well as previous rotation curve (RC) data for this galaxy. FDG is an alternative gravitational model (previously known as Newtonian Fractional-Dimension Gravity - NFDG) which does not require the dark matter (DM) paradigm.

The MW is studied here with the methods of FDG and its observed rotation curves are successfully reproduced by using a variable fractional dimension $D\left (R\right)$, following previous studies of several other galaxies which were analyzed with the same methodology. An alternative dimension function $D_{m}\left(R \right)$, based on the mass-dimension field equation, was also used and yielded less accurate fits to the experimental data. 
 
In addition, we also considered possible implications of the FDG metric, based on the presence of additional weights, on the structure of Special Relativity (SR) for spacetimes with fractional dimension. One notable, but very speculative outcome of this analysis is the possibility of an effective superluminal motion in galactic regions where the space dimension is $D<3$.
\end{abstract}

\keywords{Fractional-Dimension Gravity; Newtonian Fractional-Dimension Gravity; Modified Gravity; Modified Newtonian Dynamics; Dark Matter; Galaxies; Milky Way; Superluminal Motion}
\maketitle

\section{\label{sect:intro}Introduction}

In the last few decades the Dark Matter (DM) paradigm \cite{Clifton:2011jh,CANTATA:2021asi,Garrett:2010hd,Will:2014kxa} has become the standard solution to many astrophysical and cosmological puzzles, such as galaxy rotation curves (RCs), gravitational lensing effects, galactic collisions, cosmic microwave background anisotropies, just to name a few.

While the DM solution stood unchallenged for many years and is still considered by most scientists as the gravitational scaffolding for cosmic structures, many alternative gravitational theories which do not require DM slowly gained recognition \cite{Clifton:2011jh,CANTATA:2021asi,Garrett:2010hd,Will:2014kxa}. Some examples include Modified Newtonian Dynamics (MOND) \cite{Milgrom:1983ca,Milgrom:1983pn,Milgrom:1983zz,Famaey:2011kh}, Conformal Gravity (CG) \cite{Mannheim:1988dj,Mannheim:1992tr,Mannheim:2005bfa}, Modified Gravity (MOG) \cite{Moffat:2005si}, Fractional Gravity \cite{Calcagni:2011sz,Calcagni:2009kc,Calcagni:2010bj,Calcagni:2013yqa,Calcagni:2020ads,Calcagni:2021ipd,Calcagni:2021aap,Calcagni:2021mmj}, and several other gravitational models.

During 2020--2021, Newtonian Fractional-Dimension Gravity (NFDG) was first introduced \cite{Varieschi:2020ioh,Varieschi:2020dnd,Varieschi:2020hvp} with a modest goal of exploring classical gravitational theories for astrophysical structures characterized by a fractional-dimension  $D \lesssim 3$ (a Hausdorff-type fractal dimension). This approach was based on the increased importance of fractal geometry, as originally introduced by Mandelbrot and others \cite{1983whf..book.....M,Falconer_1985,barnsley1993fractals}, and by its application to astrophysical objects \cite{Baryshev:2002tn,Nottale:2011zz,Calcagni:2016azd}. While the overall space dimension of the universe remains the standard $D=3$ value, structures like galaxies, cluster of galaxies, etc. might be better described by a fractional-dimension function $D\left (R\right)$, usually depending on the main radial coordinate of the structure being studied and with real positive values, typically $1 <D \lesssim 3$.

NFDG was later expanded \cite{Varieschi:2022mid,Varieschi:2022nte,Varieschi:2025nzd} to include the analysis of several galaxies from the SPARC database \cite{Lelli:2016zqa,Li:2018tdo,Li:2022zms}, without using any DM components. It was also applied to the case of galaxies with little or no dark matter, such as the ultra-diffuse galaxies AGC 114905 and NGC 1052-DF2, and to the Bullet Cluster merger (1E0657-56) \cite{Varieschi:2022nte}. A more foundational NFDG equation (mass-dimension field equation) was also in introduced in Refs. \cite{Varieschi:2022mid,Varieschi:2022nte,Varieschi:2025nzd}, for a direct derivation of the variable dimension function.

In recent years, more observational evidence has accumulated regarding dark-matter-deficient (DMD) galaxies (see for example Refs. \cite{Wang:2025xre,Trivedi:2026biw,Moreno:2022wom}). Within the FDG framework, the distinction between a DM-deficient galaxy and an ordinary one is obviously based on the value of the fractal dimension $D$, mapping the spatial distribution of its baryonic matter \cite{Varieschi:2022nte}: galaxies whose $D$ values are closely aligned with standard three-dimensional space are also observed to have little or no DM in standard non-FDG analyses.

Ultimately, what distinguishes a DMD galaxy from an ordinary one in FDG is not the presence or absence of a physical substance, but rather how structurally ``fractal'' the galaxy is. This could also depend on the overall shape of the galaxy, its baryonic density, and detailed mass distributions. The mass-dimension field equation introduced in Refs. \cite{Varieschi:2022mid,Varieschi:2022nte,Varieschi:2025nzd} represents the first attempt in FDG to derive the fractal dimension $D$ directly from the baryonic mass distribution of a galaxy and is currently the only mechanism used in FDG to derive the fractal structure of a galaxy.

A relativistic version of the model was also tentatively established in another publication \cite{Varieschi:2021rzk}, which was named Relativistic Fractional-Dimension Gravity (RFDG), thus showing that also General Relativity (GR) possibly admits a fractional extension. The status of all these investigations is also available in a dedicated website \cite{Varieschi:webpage}, which is continuously updated and contains current analyses of all the galaxies studied with NFDG methods. 

Since in this paper we will combine elements of both Newtonian, non-relativistic, and relativistic versions, we have decided to simplify the name of our model. From now on, we will simply call it \textit{Fractional-Dimension Gravity} (FDG), thus including both versions into a single model.

FDG is not the only model in the literature to employ a fractional/fractal approach to astrophysics and cosmology. In addition to the already mentioned Fractional Gravity by Calcagni \cite{Calcagni:2011sz,Calcagni:2009kc,Calcagni:2010bj,Calcagni:2013yqa,Calcagni:2020ads,Calcagni:2021ipd,Calcagni:2021aap,Calcagni:2021mmj}, these techniques are also present in works by Giusti et al. \cite{Giusti:2020rul,Giusti:2020kcv}, Benetti et al. \cite{Benetti:2023nrp,Benetti:2023vxy,Benetti:2023imt}, Llanes-Estrada \cite{Llanes-Estrada:2021hnt}, Moradpour et al. \cite{Moradpour:2024uqa}, and in the $\kappa$-model by G. Pascoli \cite{Pascoli:2023dwn,Pascoli:2024med,Pascoli:2024dqg}, just to cite a few. Some of these studies are also based on fractional calculus \cite{Herrmann:2011zza,MR1890104,Varieschi:2017xjc}, and show connections with our FDG formalism.

In Section \ref{sect:potentials} we will review the FDG gravitational potentials and related applications to astrophysical structures with spherical or axial symmetries. In Section \ref{sect:milkyway}, we will model the Milky Way galaxy by considering only baryonic mass distributions, while avoiding any DM components. In Section \ref{sect:galactic}, we will fit the main RC data for the MW, by using our FDG model. Two possible fractional dimension functions, $D\left (R\right)$ and $D_{m}\left(R \right)$, will be used as it was done for other previously studied galaxies. In Sect. \ref{sect:specialrelativity}, FDG will be connected with Einstein's Special Relativity (SR) and some interesting consequences of the FDG metric will be investigated. Final conclusions will be reported in Section \ref{sect:Conclusion}.

\section{\label{sect:potentials}FDG gravitational potentials}

The original Newtonian FDG model \cite{Varieschi:2020ioh,Varieschi:2020dnd,Varieschi:2020hvp} was based on the existing mathematical theory of spaces with non-integer dimension  \cite{doi:10.1063/1.523395,1987JPhA...20.3861S,Palmer_2004}, supplemented by dimensional regularization techniques commonly used in quantum field theories \cite{Bollini:1972ui,tHooft:1972tcz,Wilson:1972cf,1995iqft.book.....P}. It was also related to Weyl's fractional integrals \cite{Varieschi:2020ioh}, thus establishing a direct connection between fractal space--time and fractional calculus.

In addition, the NFDG gravitational potentials were shown to satisfy a generalized fractional Poisson equation (Eq. (22) in Ref. \cite{Varieschi:2020ioh}), also related to a generalized D-dimensional Laplace operator. Solutions to the D-dimensional Laplace equation in spherical coordinates and corresponding multipole expansions were also presented in Ref. \cite{Varieschi:2020ioh}, as well as general theorems for spherically symmetric mass distributions in D-dimensional spaces (in Appendix A and B of the same Ref. \cite{Varieschi:2020ioh}). 

All this led to an extension of Gauss's law for the Newtonian gravitational force, which included $D +1$ spacetimes, where $D \leq 3$  can be considered a non-integer space dimension \cite{Varieschi:2020ioh,Varieschi:2020dnd,Varieschi:2020hvp}. The FDG gravitational potential for a point mass $m$ placed at the origin takes the following form \cite{Varieschi:2022mid}:
\begin{gather}\Phi _{NFDG}(r) = -\frac{2\pi ^{1 -D/2}\Gamma (D/2)\ Gm}{\left (D -2\right)l_{0}r^{D -2}}\ ;\ D \neq 2 \label{eq2.1} \\
	\Phi _{NFDG}\left (r\right) =\frac{2\ Gm}{l_{0}}\ln r\ ;\ D =2 , \nonumber \end{gather}
where $G$ is the gravitational constant. The radial coordinate $r$ is dimensionless, while the dimensional correctness of the above formulas is ensured by a scale length $l_{0}$, usually present in most fractional gravity models.

In our previous papers, the scale length $l_{0}$ was also connected to the MOND acceleration constant $a_{0} \simeq 1.2 \times 10^{ -10}\:\mbox{m}\thinspace \mbox{s}^{ -2}$ by assuming $a_{0} \approx GM/l_{0}^{2}$ for a galaxy of mass $M$. This connection became less important when it was shown that Newtonian FDG gravitational potentials and related circular velocities are independent of the choice of the $l_{0}$ parameter (see Appendix A of Ref. \cite{Varieschi:2022mid} for details). However, in Ref. \cite{Varieschi:2020ioh} it was shown that, by using the above connection between $l_{0}$ and $a_{0}$, FDG can reproduce the basic results of the MOND theory \cite{Milgrom:1983ca,Milgrom:1983pn,Milgrom:1983zz}, such as the flat rotational speed $V_{f}=\sqrt[4]{GMa_{0}}$, simply by setting $D=2$ in the FDG gravitational field.

From the expressions in Eq. (\ref{eq2.1}), Newtonian FDG formulas for axially symmetric mass distributions were derived in Ref. \cite{Varieschi:2020dnd} and then improved in Refs. \cite{Varieschi:2020hvp,Varieschi:2022mid}. In paper \cite{Varieschi:2021rzk}, we reviewed and expanded the Euler--Lagrange equations for scalar fields in fractional D-dimension spaces (rectangular, spherical, and cylindrical coordinates), together with a relativistic version of FDG, based on a modified Hilbert action.

The extension of the Newtonian FDG potentials to spherically/axially symmetric mass distributions \cite{Varieschi:2022mid} was necessary in order to model the three main components of the galactic baryonic matter: the spherical bulge mass distribution and the cylindrical stellar disk and gas distributions. These mass distributions, if not available directly from existing databases, were derived by converting existing luminosity data, such as those detailed in the SPARC database.

For this purpose, fixed light-to-mass ratios were commonly used: $\Upsilon _{disk} \simeq 0.50\ M_{ \odot }/L_{ \odot }$, $\Upsilon _{gas} \simeq 1.33\ M_{ \odot }/L_{ \odot }$ (this value included also the helium gas contribution), and $\Upsilon _{bulge} \simeq 0.70\ M_{ \odot }/L_{ \odot }$. SPARC \cite{Lelli:2016zqa,Li:2018tdo,Li:2022zms} surface luminosity data were thus turned into corresponding surface mass distributions for the three components: $\Sigma_{disk}\left (R\right)$, $\Sigma_{gas}\left (R\right)$, and $\Sigma_{bulge}\left (R\right)$. The bulge surface mass distribution could be also converted into a more suitable spherically symmetric mass distribution $\rho_{bulge} \left (r\right)$, by using Equation (1.79) in Ref. \cite{2008gady.book.....B}.

All these elements were combined into cylindrical and spherical expressions of the Newtonian FDG potentials for extended mass distributions, by using appropriate choices of coordinates and angles described in detail in Appendix A of Ref. \cite{Varieschi:2022mid}. We reproduce here the same cumbersome final expressions in order to show again the complexity of the Newtonian FDG model:

\begin{gather}\Phi _{NFDG}^{Cyl .}\left (R\right ) = -\frac{4\sqrt{\pi }\ \Gamma \left (D/2\right )G}{\left (D -2\right )\left [\Gamma \genfrac{(}{)}{}{}{D -1}{4}\right ]^{2}l_{0}}\sum \limits _{l =0}^{\infty }{\displaystyle\int \nolimits_{0}^{\infty }}\zeta \left (z^{ \prime }\right )dz^{ \prime }{\displaystyle\int \nolimits_{0}^{\infty }}c_{l ,D}\left (R^{ \prime } ,z^{ \prime }\right )\Sigma \left (R^{ \prime }\right )\frac{R_{ <}^{l}}{R_{ >}^{l +D -2}}R^{ \prime D -2}dR^{ \prime } \label{eq2.2} \\
	= -\frac{4\sqrt{\pi }\ \Gamma \left (D/2\right )G}{\left (D -2\right )\left [\Gamma \genfrac{(}{)}{}{}{D -1}{4}\right ]^{2}l_{0}}\sum \limits _{l =0}^{\infty }\Bigg\{\left [\frac{1}{R^{l +D -2}}{\displaystyle\int \nolimits_{0}^{R}}\zeta \left (z^{ \prime }\right )dz^{ \prime }{\displaystyle\int \nolimits_{0}^{\sqrt{R^{2} -z^{ \prime 2}}}}c_{l ,D}\left (R^{ \prime } ,z^{ \prime }\right )\Sigma \left (R^{ \prime }\right )\left (\sqrt{R^{ \prime 2} +z^{ \prime 2}}\right )^{l}R^{ \prime D -2}dR^{ \prime }\right ] \nonumber  \\
	+\left [R^{l}{\displaystyle\int \nolimits_{0}^{R}}\zeta \left (z^{ \prime }\right )dz^{ \prime }{\displaystyle\int \nolimits_{\sqrt{R^{2} -z^{ \prime 2}}}^{\infty }}c_{l ,D}\left (R^{ \prime } ,z^{ \prime }\right )\Sigma \left (R^{ \prime }\right )\frac{R^{ \prime ^{D -2}}}{\left (\sqrt{R^{ \prime ^{2}} +z^{ \prime 2}}\right )^{l +D -2}}dR^{ \prime }\right ] \nonumber  \\
	+\left [R^{l}{\displaystyle\int \nolimits_{R}^{\infty }}\zeta \left (z^{ \prime }\right )dz^{ \prime }{\displaystyle\int \nolimits_{0}^{\infty }}c_{l ,D}\left (R^{ \prime } ,z^{ \prime }\right )\Sigma \left (R^{ \prime }\right )\frac{R^{ \prime ^{D -2}}}{\left (\sqrt{R^{ \prime ^{2}} +z^{ \prime 2}}\right )^{l +D -2}}dR^{ \prime }\right ]\Bigg\} , \nonumber \end{gather}

\begin{gather}c_{l ,D}\left (R^{ \prime } ,z^{ \prime }\right ) ={\displaystyle\int \nolimits_{0}^{2\pi }}\left \vert \sin \phi ^{ \prime }\right \vert ^{\frac{D -3}{2}}\left \vert \cos \phi ^{ \prime }\right \vert ^{\frac{D -3}{2}}C_{l}^{\left (\frac{D}{2} -1\right )}\genfrac{(}{)}{}{}{R^{ \prime }\cos \phi ^{ \prime }}{\sqrt{R^{ \prime 2} +z^{ \prime 2}}}d\phi ^{ \prime } \label{eq2.3} \\
	c_{0 ,D} = -\frac{2^{\frac{5 -D}{2}}\pi ^{3/2}\sec \genfrac{[}{]}{}{}{\pi \left (1 +D\right )}{4}}{\Gamma \genfrac{(}{)}{}{}{5 -D}{4}\Gamma \genfrac{(}{)}{}{}{1 +D}{4}} ;c_{2 ,D} =\frac{2^{\frac{1 -D}{2}}\pi ^{1/2}(D -2)\left [\left (D -2\right )R^{ \prime 2} -2z^{ \prime 2}\right ]\Gamma \genfrac{(}{)}{}{}{D -1}{4}}{\left (R^{ \prime 2} +z^{ \prime 2}\right )\Gamma \genfrac{(}{)}{}{}{1 +D}{4}} ; . . . \nonumber  \\
	c_{1 ,D} =c_{3 ,D} = . . . =0 , \nonumber \end{gather}
	
\begin{gather}\Phi _{NFDG}^{Sph .}\left (r\right ) = -\frac{2\pi \Gamma \left (\frac{D}{2} -1\right )G}{\Gamma \left (\frac{D}{3}\right )\Gamma \genfrac{(}{)}{}{}{D}{6}l_{0}}\sum \limits _{l =0 ,2 ,4 , . . .}^{\infty }c_{l ,D}{\displaystyle\int \nolimits_{0}^{\infty }}\rho \left (r^{ \prime }\right )\frac{r_{ <}^{l}}{r_{ >}^{l +D -2}}r^{ \prime D -1}dr^{ \prime } \label{eq2.4} \\
	= -\frac{2\pi \Gamma \left (\frac{D}{2} -1\right )G}{\Gamma \left (\frac{D}{3}\right )\Gamma \genfrac{(}{)}{}{}{D}{6}l_{0}}\sum \limits _{l =0 ,2 ,4 , . . .}^{\infty }c_{l ,D}\left [{\displaystyle\int \nolimits_{0}^{_{r}}}\rho \left (r^{ \prime }\right )\frac{r^{ \prime l}}{r^{l +D -2}}r^{ \prime D -1}dr^{ \prime } +{\displaystyle\int \nolimits_{r}^{\infty }}\rho \left (r^{ \prime }\right )\frac{r^{l}}{r^{ \prime l +D -2}}r^{ \prime D -1}dr^{ \prime }\right ] , \nonumber \end{gather}

\begin{gather}c_{l ,D} =\int \nolimits_{0}^{\pi }\left \vert \sin \theta ^{ \prime }\right \vert ^{\frac{2D}{3} -1}\left \vert \cos \theta ^{ \prime }\right \vert ^{\frac{D}{3} -1}C_{l}^{\left (\frac{D}{2} -1\right )}\left (\cos \theta ^{ \prime }\right )d\theta ^{ \prime } \label{eq2.5} \\
	c_{0 ,D} =\frac{\pi \csc \genfrac{(}{)}{}{}{\pi D}{6}\Gamma \genfrac{(}{)}{}{}{D}{3}}{\Gamma \left (1 -\frac{D}{6}\right )\Gamma \genfrac{(}{)}{}{}{D}{2}} ;c_{2 ,D} =\frac{\pi \left (\frac{D}{3} -1\right )\csc \genfrac{(}{)}{}{}{\pi D}{6}\Gamma \left (\frac{D}{3}\right )}{\Gamma \left (1 -\frac{D}{6}\right )\Gamma \left (\frac{D}{2} -1\right )} ; . . . \nonumber  \\
	\int \nolimits_{0}^{2\pi }\left \vert \sin \phi ^{ \prime }\right \vert ^{\frac{D}{3} -1}\left \vert \cos \phi ^{ \prime }\right \vert ^{\frac{D}{3} -1}d\phi ^{ \prime } =\frac{2\pi \csc \genfrac{(}{)}{}{}{\pi D}{6}\Gamma \genfrac{(}{)}{}{}{D}{6}}{\Gamma \left (1 -\frac{D}{6}\right )\Gamma \genfrac{(}{)}{}{}{D}{3}} , \nonumber \end{gather}

\begin{equation}\Phi _{NFDG}\left (R\right ) =\Phi _{NFDG}^{Cyl .}\left (R\right ) +\Phi _{NFDG}^{Sph .}\left (R\right ) , \label{eq2.6}
\end{equation} 
where we can identify the radial coordinates ($r \equiv R$) in the galactic disk plane. 

In Eq. (\ref{eq2.2}), the vertical density function $\zeta \left (z^{ \prime }\right )$ can be set equal to a delta function $\delta \left (z^{ \prime }_{}\right )$ for thin disks, while for thick disks we normally use an exponential function $\zeta \left (z^{ \prime }\right ) =\frac{1}{2h_{z}}e^{ -z^{ \prime }/h_{z}}$, where $h_{z}$ is the vertical scale height. For the SPARC galaxies we analyzed in our previous papers, we used the relation \cite{Lelli:2016zqa,2010ApJ...716..234B} $\left (h_{z}/\ensuremath{\operatorname*{kpc}}\right ) =0.196\left (R_{d}/\ensuremath{\operatorname*{kpc}}\right )^{0.633}$, between the scale height $h_{z}$ and the radial scale length $R_{d}$ (available from SPARC data for each galaxy), properly rescaled by using dimensionless variables. However, in this current work we will need to modify slightly the $z'$ integration in Eq. (\ref{eq2.2}), as will be discussed in Sect. \ref{sect:milkyway}.

In all the equations above, the space dimension $D$ is considered a function of the field point coordinate $R$, i.e., $D=D(R)$ (or, using the mass-dimension function described later in this section, $D=D_{m}(R)$). The space derivatives of this function are usually neglected in the computation of the Newtonian FDG gravitational field:
\begin{equation}\mathbf{g}_{NFDG}\left (R\right) = -\frac{1}{l_{0}}\frac{d\Phi _{NFDG}\left (R\right)}{dR}\widehat{\mathbf{R}}, \label{eq2.7}
\end{equation}since we assume that the dimension $D$ changes slowly with the radial variable $R$.

The gravitational field is computed, in the galactic disk plane, as a function of the dimensionless radial coordinate $R$ in the same plane (the scale length $l_{0}$ is included into the definition of $R$, for dimensional correctness). Therefore, the variable dimension $D =D\left (R\right)$ characterizes each particular galaxy studied with FDG methods.
For all baryonic matter rotating in the main galactic plane, circular velocities are then obtained from Equation (\ref{eq2.7}):
\begin{equation}v_{circ}\left (R\right) =\sqrt{l_{0}R\left \vert \mathbf{g}_{NFDG}\left (R\right)\right \vert }, \label{eq2.8}
\end{equation}
with all fields computed by using the variable dimension function $D =D\left (R\right)$, or $D=D_{m}(R)$, mentioned above.

A fundamental element of the FDG computation is the series expansion of the $1/r^{D-2}$ term in the first line of Equation (\ref{eq2.1}).
Using the distance between the field point $\mathbf{x}$ and the source point $\mathbf{x}^{ \prime }$, the Euler kernel $1/\left \vert \mathbf{x} -\mathbf{x}^{ \prime }\right \vert ^{D -2}$ can be expanded for $D >1$, $D \neq 2$, with rescaled spherical coordinates \cite{2012JPhA...45n5206C}:
\begin{equation}\frac{1}{\left \vert \mathbf{x} -\mathbf{x}^{ \prime }\right \vert ^{D -2}} =\sum \limits _{l =0}^{\infty }\frac{r_{ <}^{l}}{r_{ >}^{l +D -2}}C_{l}^{\left (\frac{D}{2} -1\right)}\left (\cos \gamma \right), \label{eq2.9}
\end{equation}
where $r_{ <}$ ($r_{ >}$) is the smaller (larger) of $r$ and $r^{ \prime }$, $\gamma $ is the angle between the unit vectors $\widehat{\mathbf{r}}$ and $\widehat{\mathbf{r}}^{ \prime }$, and $C_{l}^{\left (\lambda \right)}\left (x\right)$ denotes Gegenbauer polynomials \cite{NIST:DLMF}.
Equation (\ref{eq2.9}) is then used for both spherically symmetric galactic components (bulge) and cylindrical components (disk and gas), with appropriate choices of coordinates and angles as described in detail in Ref. \cite{Varieschi:2022mid}. After further mathematical elaboration, the final Newtonian FDG potentials in Eqs. (\ref{eq2.2}) and (\ref{eq2.4}) are obtained, where the summations over the $l$ index in both equations are due to the series expansion in Eq. (\ref{eq2.9}).

These summations, based on the Euler kernel, are usually rapidly converging, thus allowing for the inclusion of only the first few terms. In this work, we will include the first six non-zero terms of the expansions  ($l =0,2,4,6,8,10$, in both Eqs. (\ref{eq2.2}) and (\ref{eq2.4})), by using also the $c_{l ,D}$ coefficients outlined in Eqs. (\ref{eq2.3}) and (\ref{eq2.5}).

As discussed at length in our last paper \cite{Varieschi:2025nzd}, the Newtonian FDG potentials/fields/circular velocities are complicated expressions which are not suitable for a direct fit to the galactic experimental data. Therefore, we will continue using two different strategies for the fitting of the galaxy rotation data.

The first strategy is to find the fractional dimension function $D\left (R\right)$ by matching directly the experimental data with the Newtonian FDG formulas for a fixed number of points, over the range of radial distances. This approach usually produces a perfect fit to the galactic data, since we are simply computing, at each radial location, the value of $D(R)$ which matches the experimental data.

In our latest paper \cite{Varieschi:2025nzd}, a more foundational approach was introduced by using a so-called mass-dimension field equation:
 \begin{equation} D_{m}\left(R \right)-1 +R\ln\left(\frac{R}{R_{0}} \right)\frac{dD_{m}\left(R \right)}{dR}   = \frac{2\pi^{\left(\frac{D_{m}\left(R \right)-1}{2} \right)}}{\Gamma\left(\frac{D_{m}\left(R \right)-1}{2}  \right)} \frac{\tilde{\Sigma}_{tot}\left({R}/{R_{0}} \right)}{M_{0}}, \label{eq2.10}
 \end{equation}
which can be solved numerically for the mass-dimension function $D_m(R)$.

This new dimension function $D_m(R)$ plays the same role of the previous $D(R)$ function, but is obtained solely from the known distribution of baryonic matter in the galaxy being studied, and without any DM~contributions. 
In the previous equation, the quantity $R_{0}$ acts as a scale length for our fractional-dimension structures, equivalent to the original FDG scale length $l_{0}$ in Eq. (\ref{eq2.1}). The total surface mass density $\tilde{\Sigma}_{tot}(R/R_{0})$ in the last equation, is a ``rescaled'' mass density, i.e., $\tilde{\Sigma}_{tot}(R/R_{0})=\Sigma_{tot}(R) \: R_{0}^{2}$, with dimensions of mass.

This mass distribution, used in Eq. (\ref{eq2.10}), is based on the total surface mass distribution $\Sigma_{tot} \left (R\right) =\Sigma _{bulge}\left (R\right) +\Sigma _{disk}\left (R\right)+\Sigma _{gas}\left (R\right)$, for the three main components. The bulge mass distribution, usually described by a spherical distribution $\rho_{bulge} \left (r\right)$, is here given in terms of a surface mass density $\Sigma_{bulge}\left (R\right)$, since it is usually obtained from a surface bulge luminosity. Therefore, all the three mass distributions are treated as axially symmetric and the total surface mass distribution is used in Eq. (\ref{eq2.10}).

Full derivation of the mass-dimension field equation (\ref{eq2.10}) can be found in Ref. \cite{Varieschi:2025nzd}. In the following two sections, we will first model the mass distribution functions for the Milky Way galaxy and then apply FDG methods to fit the related RCs.

\section{\label{sect:milkyway}FDG modeling of the Milky Way galaxy}

Following the discussion in the previous section, the main input of FDG is the baryonic mass distribution of the galaxy being studied, divided into the three fundamental components: spherical bulge, stellar and gas disk distributions, without any DM halo, or DM disk contributions. Then, assuming a precise form for the variable dimension function $D(R)$ or $D_m(R)$, the circular velocity data for the galaxy can be fitted by using Eqs. (\ref{eq2.2})-(\ref{eq2.8}) and the full RC is finally obtained.

General reviews of the MW rotation curve and related mass models can be found in works by Bhattacharjee et al. \cite{Bhattacharjee:2013exa}, Sofue \cite{Sofue:2020rnl} (and references therein), but we will use here the recent paper by Sylos Labini et al. \cite{Labini:2023fmy} as the main reference for our analysis. All current papers on the MW rotation curves use data from the Gaia Mission Data Release 3 (Gaia DR3) \cite{refId0}, since these recent data yield rotation curves showing a definite decline of the RC at larger distances  \cite{Labini:2023fmy}, as opposed to previous RCs \cite{Sofue:2020rnl} where this decline was less noticeable.

In this work, we will use two main rotation curves for the Milky Way galaxy: the unified RC from the galactic center to the radial distance of about $100~kpc$ compiled by Sofue (data in Tables A1-A2 of Ref. \cite{Sofue:2020rnl}), and the more recent RC compiled by Sylos Labini et al. (Table 1 in Ref. \cite{Labini:2023fmy}), based on Gaia DR3 data. This latter RC combines analyses of the Gaia DR3 data by Eilers et al. \cite{Eilers:2019gqs} and Wang et al. \cite{2023ApJ...942...12W}, but is more limited in the range of galactocentric radial distances (about $5-28~kpc$). Other RCs for the MW were also introduced in more recent papers (see \cite{2024ApJ...976..185S,Feng:2026huz} and references therein), but we will limit our analysis to the two rotation curves mentioned above.
 
\begin{figure}\centering 
	\setlength\fboxrule{0in}\setlength\fboxsep{0.1in}\fcolorbox[HTML]{000000}{FFFFFF}{\includegraphics[width=6.80in, height=8.00in,]{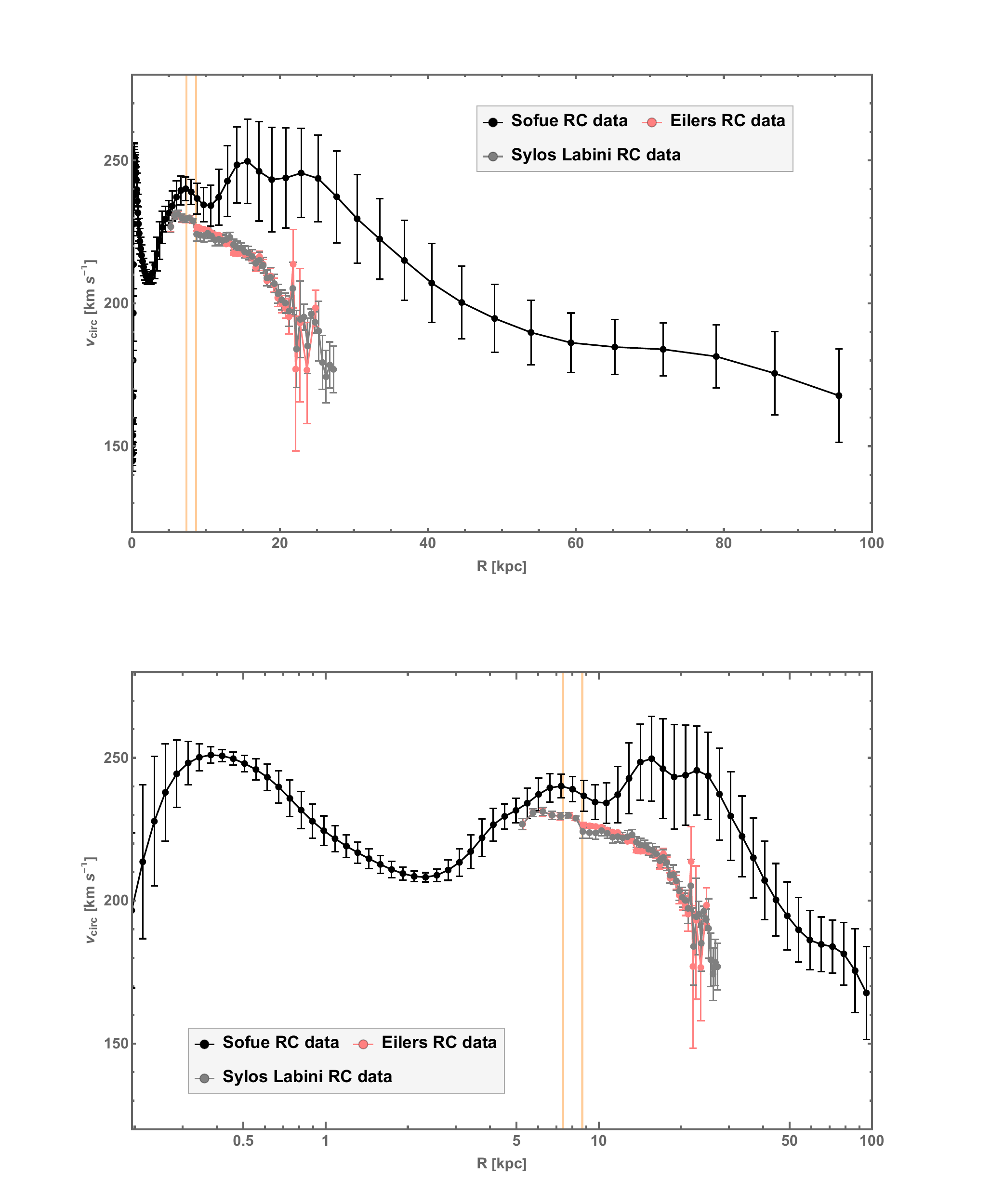}}
	\caption{Milky Way rotation curves.
		Top panel: Sofue \cite{Sofue:2020rnl} unified RC data (black circles with error bars) and continuous line interpolating the data points; Sylos Labini et al. \cite{Labini:2023fmy} RC data (gray circles with error bars) with related interpolating line; Eilers et al. \cite{Eilers:2019gqs} RC data (light red circles with error bars) with related interpolating line. Linear scale for the galactocentric distance $R$, over the range $0-100~kpc$. Bottom panel: same as the top panel, but with a logarithmic scale for the radial distance $R$.
		Also shown in both panels: approximate radial position of the Solar System (vertical orange lines) in the range $7.4–8.7~kpc$.}
	\label{figure:MWfig1}
\end{figure}

In Fig. \ref{figure:MWfig1} we show these main MW rotation curves: Sofue's data in black and Sylos Labini's data in gray, both with related error bars and continuous lines interpolating the data sets. We also show, for comparison, RC data from Eilers et al. \cite{Eilers:2019gqs} (light red), but we will not use these data in our subsequent analysis since they are very close to those compiled by Sylos Labini. We present the same data in two different panels: top panel with a linear scale for the galactocentric radial distance, bottom panel with a logarithmic scale for the same variable $R$.
As already mentioned, the more recent RCs (Sylos Labini, Eilers, etc.) show a more prominent decline of the circular velocities, but are affected by greater uncertainties at larger distances. 

In both panels, we also show with vertical orange lines the approximate radial position of the Solar System, whose distance from the galactic center is estimated to be within the range $24–28.4$ \textrm{kilo-light-years} ($7.4–8.7~kpc$) \cite{Francis:2013zna}.
While the goal of our analysis is to fit the experimental data in Fig. \ref{figure:MWfig1}, the main input of our FDG computations is the baryonic mass distribution of the galactic bulge, stellar disk, and gas components. Among the many possible models in the literature, we have chosen the following mass distributions, based on \cite{Labini:2023fmy} and references therein.

The stellar components will consist of the spherical bulge and thin/thick disks, following Eilers et al. \cite{Eilers:2019gqs}. Eilers used Miyamoto-Nagai profiles \cite{Miyamoto:1975zz} for the thin/thick disks and a spherical Plummer potential \cite{Plummer:1911zza} for the bulge. These formulas are shown in Eq. (\ref{eq3.1}), respectively equivalent to Eq. (2.69b) and Eq. (2.44b) in Binney-Tremaine \cite{2008gady.book.....B}, yielding the following mass distributions:
\begin{gather}\rho_{M}\left( R,z \right)=\left( \frac{b^{2}M}{4\pi} \right)\frac{aR^{2}+\left(a+3\sqrt{z^{2}+b^{2}}  \right)\left( a+\sqrt{z^{2}+b^{2}} \right)^{2}}{\left[ R^{2}+\left( a+\sqrt{z^{2}+b^{2}} \right)^{2} \right]^{5/2}\left( z^{2}+b^{2} \right)^{3/2}} \label{eq3.1} \\
	\rho_{P}\left( r \right)=\frac{3M}{4\pi b^{3}}\left( 1+\frac{r^{2}}{b^{2}} \right)^{-5/2}. \nonumber \end{gather}

In addition, parameters from model I in Table 1 of Pouliasis et al. \cite{2017A&A...598A..66P} were used. For the Miyamoto-Nagai density profile $\rho_{M}$ in the first line of Eq. (\ref{eq3.1}): $M_{thin}=1700.0$, $M_{thick}=1700.0$, $a_{thin}=5.3000$, $a_{thick}=2.6$, $b_{thin}=0.25$, $b_{thick}=0.8$. For the Plummer spherical density $\rho_{P}$ in the second line of Eq. (\ref{eq3.1}): $M_{bulge}=460.0$, $b_{bulge}=0.3000$, where all these masses are in units of $2.32\times 10^{7} M_{\odot}$ and all distances in $kpc$.

In the spherical bulge distribution, we also added the presence of a massive black hole, near the galactic center of the MW, of mass $M_{BH} \simeq 4.3 \times 10^{6} M_{\odot}$ and approximate radius $r_{BH} \simeq 25.9 \times 10^{6}~km$, corresponding to the Sagittarius $A^{*}$ black hole \cite{2022ApJ...930L..12E,2023A&A...677L..10G}, and modeled as:
\begin{equation}\rho_{BH}\left( r \right)=\frac{3 M_{BH}}{4\pi r_{BH}^{3}}\left[\mathcal{H}(r)-\mathcal{H}(r-r_{BH}) \right], \label{eq3.2} \end{equation} 
where $\mathcal{H}(x)$ indicates the Heaviside step function.

Using Eqs. (\ref{eq3.1})-(\ref{eq3.2}) and all the parameters above, the stellar disk distribution will be the combination of the thin and thick disk components with Miyamoto-Nagai density profiles in Eq. (\ref{eq3.1}), while the bulge spherical distribution will combine the Plummer spherical density in Eq. (\ref{eq3.1}) with the black hole spherical density in Eq. (\ref{eq3.2}):
\begin{gather}\rho_{disk}\left( R,z \right)=\rho_{M,thin}\left( R,z \right)+\rho_{M,thick}\left( R,z \right) \label{eq3.3} \\
	\rho_{bulge}\left( r \right)=\rho_{P}\left( r \right)+\rho_{BH}\left( r \right). \nonumber \end{gather}

For the gas distribution of atomic $H\thinspace\textsc{i}$ and molecular $H_{2}$, we follow again Sylos Labini et al. \cite{Labini:2023fmy} and parameterize the surface mass distribution as:
\begin{equation}\Sigma_{H\thinspace\textsc{i}+H_{2}}\left( R \right)=\frac{\Sigma_{0}}{\left[1+\left(\frac{R}{R_{d}}\right)^{\alpha}\right]\left(\frac{R}{R_{d}}\right)^{0.25}}, \label{eq3.4} \end{equation} 
with $\Sigma_{0} \approx 6 \times 10^{6}~M_\odot/kpc^{2}$, $R_{d}=17~kpc$, and $\alpha=10$. This formula provides the best analytical fit to the observed surface density data for atomic and molecular hydrogen in the MW \cite{Bigiel:2012wj}. By numerically integrating all these mass distributions over the galactic volume, we estimated the total baryonic mass of the MW as $M_{total} \simeq 9.61 \times 10^{10} M_{\odot}=1.91 \times 10^{41}~kg$, in line with existing values in the literature.

The mass distributions in Eqs. (\ref{eq3.1})-(\ref{eq3.4}) become the main input of our FDG Eqs. (\ref{eq2.2})-(\ref{eq2.6}), and of the mass-dimension field equation (\ref{eq2.10}). However, some modifications are needed in view of the MW mass distribution functions described above in Eqs. (\ref{eq3.1})-(\ref{eq3.4}). The thin/thick disk mass distribution $\rho_{disk}\left( R',z' \right)$ in the first line of Eq. (\ref{eq3.3}), with the Miyamoto-Nagai profiles in the first line of Eq. (\ref{eq3.1}), is entered into Eq. (\ref{eq2.2}) in place of the combined functions $\zeta \left (z^{ \prime }\right )\Sigma \left (R^{ \prime }\right )$ and all the double integrals are carried out accordingly, in our Mathematica routines.

The gas surface mass distribution $\Sigma_{H\thinspace\textsc{i}+H_{2}}\left( R' \right)$ in Eq. (\ref{eq3.4}) can be entered directly into Eq. (\ref{eq2.2}) in place of the $\Sigma \left (R^{ \prime }\right )$ function, and the vertical $z'$ integration is carried out as discussed in the paragraph immediately following Eq. (\ref{eq2.6}). The bulge mass distribution $\rho_{bulge}\left( r' \right)$ in the second line of Eq. (\ref{eq3.3}) can be entered directly into Eq. (\ref{eq2.4}) in place of the $\rho \left (r^{ \prime }\right )$ function, without any additional modifications.

The Newtonian FDG gravitational potential is now the sum of the potentials obtained by using the three mass components (thin/thick disk, gas, bulge), so we modify Eq. (\ref{eq2.6}) as:
\begin{equation}\Phi _{NFDG}\left (R\right ) =\Phi _{NFDG}^{disk}\left (R\right ) +\Phi _{NFDG}^{gas}\left (R\right )  +\Phi _{NFDG}^{bulge}\left (R\right ), \label{eq3.5}
\end{equation}where, again, we can identify the radial coordinates ($r \equiv R$) in the galactic disk plane. The computation of the Newtonian FDG field and circular velocities then proceeds according to Eqs. (\ref{eq2.7})-(\ref{eq2.8}) in Sect. \ref{sect:potentials}.

Similar procedures were used to convert the thin/thick disk and bulge distributions into equivalent surface mass distributions. These were added to the gas surface mass distribution of Eq. (\ref{eq3.4}), and the total rescaled surface mass distribution was used as $\tilde{\Sigma}_{tot}(R/R_{0})$ inside the mass-dimension field equation (\ref{eq2.10}). This procedure allowed for the alternative computation of the mass-dimension function $D_{m}(R)$, directly from the baryonic masses. Once again, we stress that no dark matter components were ever used in our FDG computations.

\section{\label{sect:galactic}FDG galactic data fitting}

In this section, we will apply the FDG methods outlined in Sects. \ref{sect:potentials}-\ref{sect:milkyway} and fit the MW circular velocity data, without using any DM. As previously mentioned, we will fit the Sylos Labini et al. RC data, over the more limited range of radial distances ($R \approx 5-28~kpc$), and the Sofue RC data over the extended range $R \approx 0-100~kpc$. We will not fit the Eilers et al. data included in Fig. \ref{figure:MWfig1}, since they are very similar to the Sylos Labini et al. data.

\begin{figure}\centering 
	\setlength\fboxrule{0in}\setlength\fboxsep{0.1in}\fcolorbox[HTML]{000000}{FFFFFF}{\includegraphics[width=6.80in, height=8.00in,]{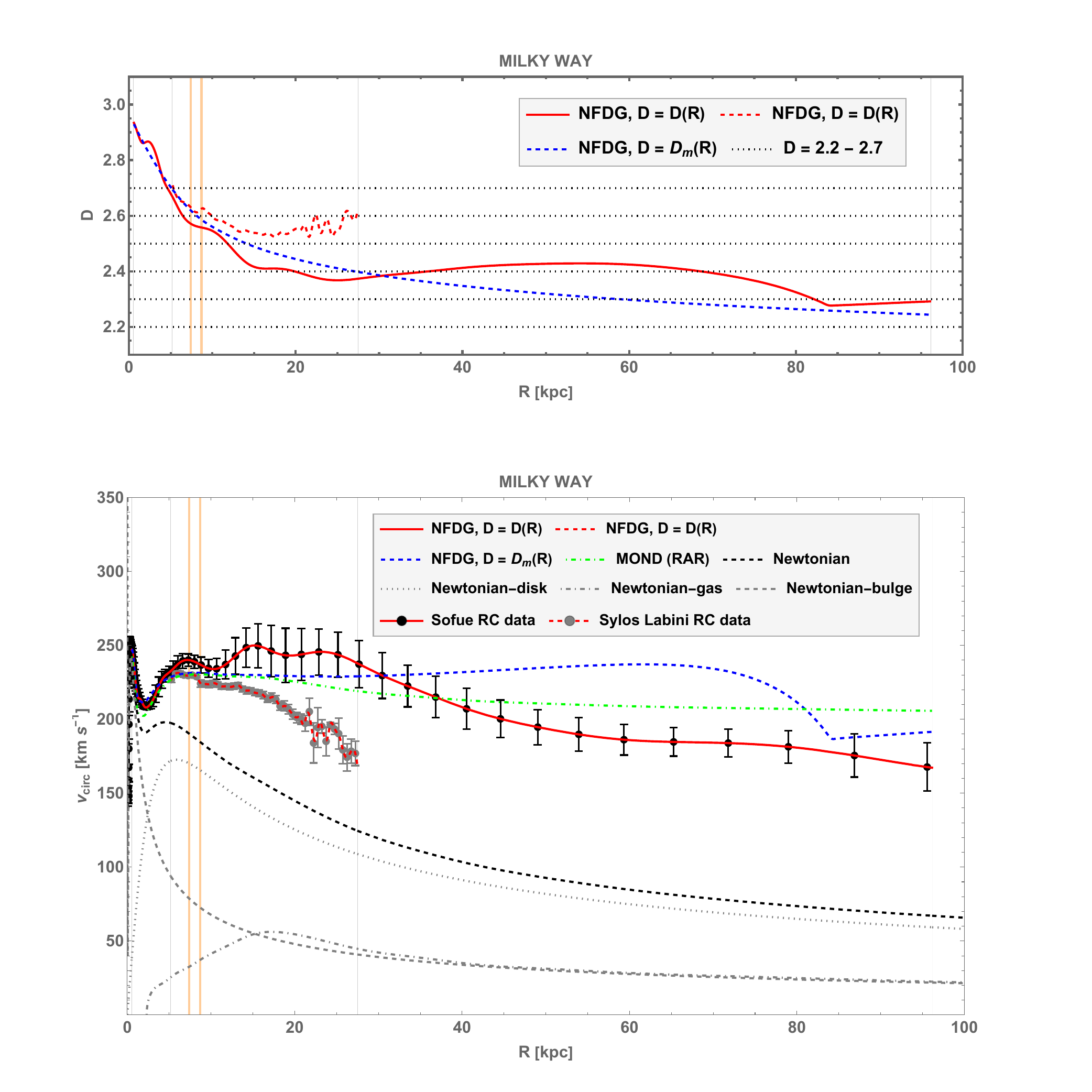}}
	\caption{Newtonian FDG results for Milky Way galaxy, with linear radial distances.
		Top panel: NFDG variable dimension $D\left (R\right )$, based directly on Sofue data (red-solid curve) and on Sylos Labini data (red-dashed curve), compared with NFDG mass-dimension $D_{m}\left(R\right)$ (blue-dashed curve), and fixed values $D =2.2-2.7$ (black-dotted lines). Bottom panel: NFDG rotation curves (circular velocity vs. radial distance) compared to the original data (black circles with error bars for Sofue data, gray circles with error bars for Sylos Labini data). The NFDG best fits for the variable dimension $D\left (R\right )$ are shown by the red-solid/red-dashed lines, while the NFDG fit for the mass-dimension $D_{m}\left(R\right)$ is shown by the blue-dashed curve. Also shown: MOND prediction based on the general RAR (green, dot-dashed), Newtonian rotation curves (total - black dashed line, different components - gray lines), and approximate radial position of the Solar System (vertical orange lines).}
	\label{figure:MWfig2}
\end{figure}

\begin{figure}\centering 
	\setlength\fboxrule{0in}\setlength\fboxsep{0.1in}\fcolorbox[HTML]{000000}{FFFFFF}{\includegraphics[ width=6.80in, height=8.00in,]{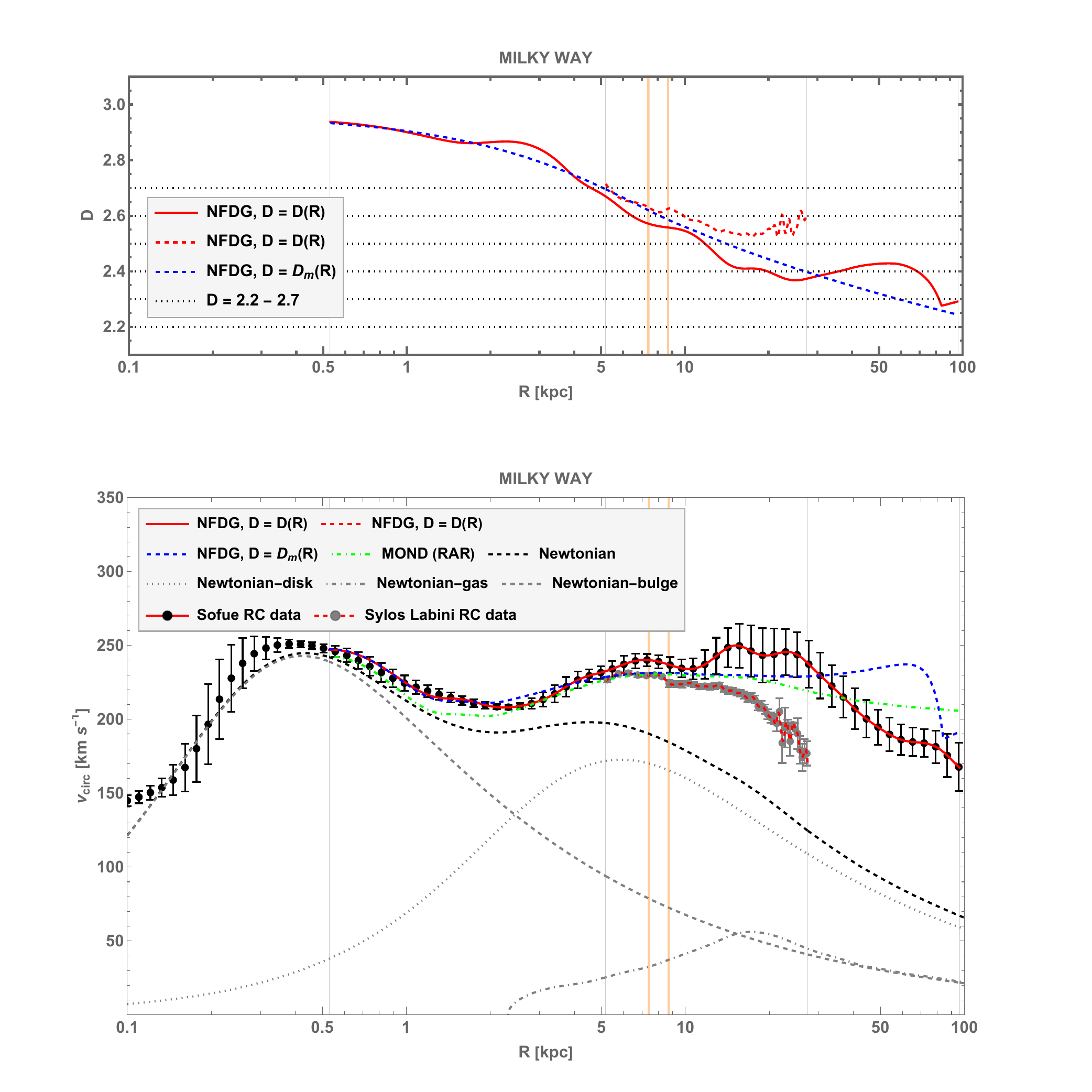}
	}
	\caption{Newtonian FDG results for Milky Way galaxy, with logarithmic radial distances.
		Top panel: NFDG variable dimension $D\left (R\right )$, based directly on Sofue data (red-solid curve) and on Sylos Labini data (red-dashed curve), compared with NFDG mass-dimension $D_{m}\left(R\right)$ (blue-dashed curve), and fixed values $D =2.2-2.7$ (black-dotted lines). Bottom panel: NFDG rotation curves (circular velocity vs. radial distance) compared to the original data (black circles with error bars for Sofue data, gray circles with error bars for Sylos Labini data). The NFDG best fits for the variable dimension $D\left (R\right )$ are shown by the red-solid/red-dashed lines, while the NFDG fit for the mass-dimension $D_{m}\left(R\right)$ is shown by the blue-dashed curve. Also shown: MOND prediction based on the general RAR (green, dot-dashed), Newtonian rotation curves (total - black dashed line, different components - gray lines), and approximate radial position of the Solar System (vertical orange lines).}
	\label{figure:MWfig3}
\end{figure}

The FDG results are shown in Fig. \ref{figure:MWfig2} and Fig. \ref{figure:MWfig3}. The only difference between these two figures is the scale used for the radial distance $R$: linear in Fig. \ref{figure:MWfig2}, logarithmic in Fig. \ref{figure:MWfig3}. The linear scale allows for a better view of the results at larger distances, while the logarithmic scale is more effective at lower radial distances.

As done in past papers, we will measure the radial distance $R$ in \textrm{kiloparsec} and the circular velocities $v_{circ}$ in $\mbox{km}\ \mbox{s}^{ -1}$. The computations are based on the methods detailed in Sections \ref{sect:potentials}--\ref{sect:milkyway}, with the radial limits for our data fitting procedures set at $R_{\min } =5.2\ \mbox{kpc}
$ and $R_{\max } =27.5\ \mbox{kpc}$ for the Sylos Labini data, and $R_{\min } =0.53\ \mbox{kpc}
$ and $R_{\max } =96.1\ \mbox{kpc}$ for the Sofue data. These limits are shown in both figures as vertical thin-gray lines.

The main FDG result, in both Figs. \ref{figure:MWfig2}-\ref{figure:MWfig3}, is the variable dimension $D\left (R\right)$ shown in the top panel by the red-solid/red-dashed curves, following the Sofue's and Sylos Labini's RC data, respectively. These  $D\left (R\right)$ curves were obtained by interpolating the experimental data for the circular velocities (black/gray circles with error bars in the bottom panel) and by calculating the observed gravitational field $g_{obs}\left (R\right)$, based solely on these experimental data. This observed field was then assumed to be the same as $g_{NFDG}\left (R\right)$ in Equation (\ref{eq2.7}), derived from the Newtonian FDG gravitational potentials discussed in Section \ref{sect:potentials}.

The Newtonian FDG potential and related gravitational field at each field point, are considered to be functions of the variable dimension $D\left (R\right)$, i.e., $g_{NFDG}\left (R,D\left (R\right)\right)$. The computational range $\left (R_{\min },R_{\max }\right)$ for each set of data was divided into equal sub-intervals with related radial distances $R_{i} =R_{\min } +i\genfrac{(}{)}{}{}{R_{\max } -R_{\min }}{n_{data}},\ \ i =0, \dots,n_{data}$, with $n_{data}=50$ for Sylos Labini's data and $n_{data}=200$ for Sofue's data. 

We then solved numerically the following equation:
\begin{equation}g_{NFDG}\left (R_{i},D\left (R_{i}\right)\right) =g_{obs}\left (R_{i}\right), \label{eq4.1}
\end{equation}
for each of the $R_{i}$ points, related to the two different data sets, thus leading to the corresponding values for the variable dimension $D_{i} \equiv D\left (R_{i}\right),\ i =1, \dots,n_{data}$.

The NFDG red-solid/red-dashed $D\left (R\right)$ curves in the top panels of Figs. \ref{figure:MWfig2}-\ref{figure:MWfig3} are obtained by interpolating these sets of $\left (R_{i},D_{i}\right)$ points. To double-check our procedure, we then recomputed the NFDG circular velocities using the $D\left (R\right)$ functions and Eq. (\ref{eq2.8}), at each radial point $R_{i}$. This produces the  final NFDG fits to the experimental data shown by the red-solid/red-dashed curves in the bottom panels of the same figures. The perfect agreement of these fits with the experimental data is expected. At each point, we have chosen the appropriate value of the variable dimension $D\left (R_{i}\right)$ that yields a perfect match between the experimental $g_{obs}\left (R_{i}\right)$ and the predicted NFDG field $g_{NFDG}\left (R_{i},D\left (R_{i}\right)\right)$, in view of Eq. (\ref{eq4.1}).

An alternative method, introduced in our paper \cite{Varieschi:2025nzd} and based on Eq. (\ref{eq2.10}), can be used to derive the NFDG mass-dimension function $D_{m}\left(R\right)$ directly from the bulge/disk/gas mass distributions of the MW. This method yields the blue-dashed curve in the top panels for $D_{m}\left(R\right)$, and the corresponding blue-dashed curves for the circular velocity in the bottom panels of Figs. \ref{figure:MWfig2}-\ref{figure:MWfig3}. This method was only used for the analysis of the extended RC (Sofue's data) and not for the limited-range RC (Sylos Labini's data).

For this analysis, a ``NonlinearModelFit'' Mathematica function was used in our routines, with the model defined as the differential equation (\ref{eq2.10}). Free parameters $R_{0}$ and $M_{0}$ were set in the model, together with an initial condition $D_{m}\left(R_{data} \right)=D\left(R_{data} \right)$. This initial condition was set by matching one of the last Sofue's data points for this galaxy, at a distance $R_{data}$, and whose circular velocity is close to the asymptotically flat rotation velocity for the MW, estimated as $V_{data}\approx 230~km/s$.

The outcome of this procedure is represented by the blue-dashed curves in both panels of Figs. \ref{figure:MWfig2}-\ref{figure:MWfig3}. As remarked in previous analyses of other galaxies, the blue-dashed fits to the circular velocity data in the bottom panels are not as accurate as those described by the red-solid curves. Nevertheless, they are able to capture the overall trend of the data, especially at lower radial distances, as can be better seen in Fig. \ref{figure:MWfig3}, using the logarithmic distance scale. Similarly, the blue-dashed $D_{m}\left(R\right)$ curves in the top panels are a good approximation for the variable dimension, but do not show the details of the red-solid $D\left(R\right)$ functions.

Therefore, the field equation (\ref{eq2.10}) for $D_{m}\left(R\right)$ is effective in predicting the rotational velocity general pattern of the MW galaxy without any DM contributions, but is not as accurate as the method based on the original $D\left (R\right)$. However, both methods confirm the validity of our FDG approach to the MW dynamics.

In the bottom panels of Figs. \ref{figure:MWfig2}-\ref{figure:MWfig3}, we also show the MOND curves (green, dot-dashed) based on the general Radial Acceleration Relation (RAR) formula \cite{McGaugh:2016leg,Lelli:2017vgz}:
\begin{equation}g_{obs} =\frac{g_{bar}}{1 -e^{ -\sqrt{g_{bar}/g_{\dag }}}}, \label{eq4.2}
\end{equation}
where $g_{\dag } =1.20 \times 10^{ -10}\ \mbox{}\ \mbox{m}\thinspace \mbox{s}^{ -2}$ is equivalent to the MOND acceleration scale $a_{0}$.

This MOND (RAR) fit was obtained by using directly Eq. (\ref{eq4.2}) and the Sofue's data for $g_{bar}$, without further adjustments, as opposed to usual MOND analyses where additional quantities are used as free parameters to improve the fitting procedure. These MOND (RAR) fits, without adjustment of the parameters, are only able to approximately describe the overall pattern of the empirical data, without yielding good fits to the data. In the bottom panels of Figs. \ref{figure:MWfig2}-\ref{figure:MWfig3}, we also show the Newtonian rotation curves (different components-gray lines; total-black dashed line). These Newtonian curves are obtained from direct interpolation of the experimental data.

Going back to the main results, represented by the red-solid/red-dashed variable dimension curves in the top panels of both Figs. \ref{figure:MWfig2} and \ref{figure:MWfig3}, the FDG interpretation is the same as for all previously analyzed galaxies. If we postulate that the MW galaxy behaves as a fractal structure whose Hausdorff fractional dimension follows the $D\left (R\right)$ functions, then the red-solid/red-dashed rotation curves will match the experimental data without needing any form of DM.

The $D\left (R\right)$ curves seem to be increasing toward standard $D \approx 3$ values at lower radial distances where the spherical bulge is more dominant, while decreasing at larger distances and approaching values of $D \approx 2.3-2.4$ for the FDG analysis of Sofue's data, or values of $D \approx 2.5-2.6$ for the FDG analysis of Sylos Labini's data. Similar behavior was observed for other galaxies studied in the past (such as NGC 5033, NGC 6674, NGC 6946, NGC 7814, \cite{Varieschi:webpage}), where the dominating stellar disk component is combined with a strong central bulge, typically yielding $D \approx 2.4-2.6$ at larger radial distances. Therefore, the MW seems to be characterized by a fractal structure similar to those of the aforementioned galaxies, in view of our FDG analysis. 

In both Figs. \ref{figure:MWfig2}-\ref{figure:MWfig3}, we also show with vertical orange lines the approximate radial position of the Solar System ($24–28.4$ \textrm{kilo-light-years}, or $7.4–8.7~kpc$ \cite{Francis:2013zna}), as was done in Fig. \ref{figure:MWfig1}. From the top panels of Figs. \ref{figure:MWfig2}-\ref{figure:MWfig3}, we can estimate the fractional dimension at the galactic radial position of the Solar System to be approximately $D_{SS} \approx 2.5-2.6$, depending on the different evaluations of $D(R)$ or $D_{m}(R)$.

This does not imply that, in the region of our Solar System, the space dimension is reduced from the standard $D=3$ to a lower $D_{SS} \approx 2.5-2.6$ value. In FDG, the local value of the space dimension is determined by the presence of local massive sources, such as stars, black holes, etc., which typically induce the standard $D=3$ space dimension.

However, just outside the Solar System and in the interstellar region between our Sun and neighboring stars, FDG assumes that the space dimension might be reduced to the lower values $D_{SS} \approx 2.5-2.6$ estimated above. Obviously, this is a very speculative hypothesis. If true, it might not just explain galactic dynamics without DM, but also lead to possible consequences for kinematics and dynamics which will be outlined in the following section.

\section{\label{sect:specialrelativity}FDG and Special Relativity}

In this section we will consider some implications of FDG for kinematics, dynamics, and Special Relativity. The contents of this section are highly speculative and not based on any direct experimental evidence; nevertheless, they should be studied as a consequence of the fractal behavior of galactic structures, as assumed in FDG.

As mentioned in the previous section, the region of our Solar System, as well as that of any other stellar system, might be characterized by a standard space dimension ($D=3$) due to the relatively high local matter density (estimated to be around $10^{-9}-10^{-10}~kg/m^{3}$), compared to the extremely low baryonic density of the MW (estimated to be of the order of  $10^{-21}-10^{-22}~kg/m^{3}$). This might be the reason why we experience standard three-dimensional gravity inside our Solar System, while the fractional dimension determined by FDG methods in the previous section, at the galactic radial distance $R\simeq8~kpc$ of the Solar System, was estimated as $D_{SS}\simeq2.5-2.6$. 

In other words, an observer traveling from the inside of the Solar System to the neighboring outer regions of our galaxy, might experience a continuous transition from local $D=3$ to lower values $D<3$ of the space dimension. Similarly, 3-dimensional gravity would be experienced in the proximity of any massive 3-dimensional object (stars, planets, black holes, etc.), but with fractal $D<3$ behavior becoming prominent away from these massive sources.

The MW fractional dimension function $D(R)$, or $D_{m}(R)$, detailed in Sect. \ref{sect:galactic}, would just be capturing the overall galactic space dimension, without showing the local transitions to standard $D=3$ in the vicinity of massive sources of gravity. The Newtonian, or almost-Newtonian behavior, near galactic centers would also be due to the increased baryonic mass density in these inner regions, the presence of super-massive black holes, etc.

Let's consider a possible transition from a $D=3$ ``standard'' frame of reference and a ``fractional" one with $D<3$. No relative motion between these frames needs to be included: the two systems can be sitting on top of each other, or be separated by a fixed distance between their origins, corresponding to the distance between the $D=3$ region and the $D<3$ region (e.g., between the Solar System $D=3$ region and the $D_{SS}\simeq2.5-2.6$ outer regions).

Considering rectangular coordinates $x_{i},~i=1,2,3$, for the standard $D=3$ reference frame, the mathematical theory of spaces with non-integer dimension \cite{doi:10.1063/1.523395,1987JPhA...20.3861S,Palmer_2004} introduces infinitesimal measures $d\mu_{i}$ for the fractional coordinates $\mu_{i},~i=1,2,3$, defined as:
\begin{equation}d\mu _{i}(x_{i}) =\frac{\pi ^{d _{i}/2}}{\Gamma (d _{i}/2)}\left \vert \frac{x_{i}}{l_{i}}\right \vert ^{d_{i} -1}dx_{i} ,\ i =1 ,2 ,3, \label{eq5.1}\end{equation}
where each coordinate can have a fractional dimension $0<d_{i}\le1$ and $D=d_{1}+d_{2}+d_{3}$.\footnote{In previous FDG papers, $\alpha_{i}$ was used to indicate the individual fractional dimensions of each coordinate. Here we prefer to use $d_{i}$ instead.} The scale lengths $l_{i}, i=1,2,3$ are similar to the scale length $l_{0}$ used in previous FDG equations, and are required for dimensional correctness. For $d_{i}=1$, the fractional measures $d\mu_{i}$ reduce to the standard coordinate differentials $dx_{i}$ and the difference between these two quantities is simply due to the presence of the ``fractional weight" $\frac{\pi ^{d_{i}/2}}{\Gamma (d_{i}/2)}\left \vert \frac{x_{i}}{l_{i}}\right \vert ^{d_{i} -1}$ (see also Ref. \cite{Varieschi:2021rzk} for more details).

\begin{figure}\centering 
	\setlength\fboxrule{0in}\setlength\fboxsep{0.1in}\fcolorbox[HTML]{000000}{FFFFFF}{\includegraphics[width=6.50in, height=4.50in,]{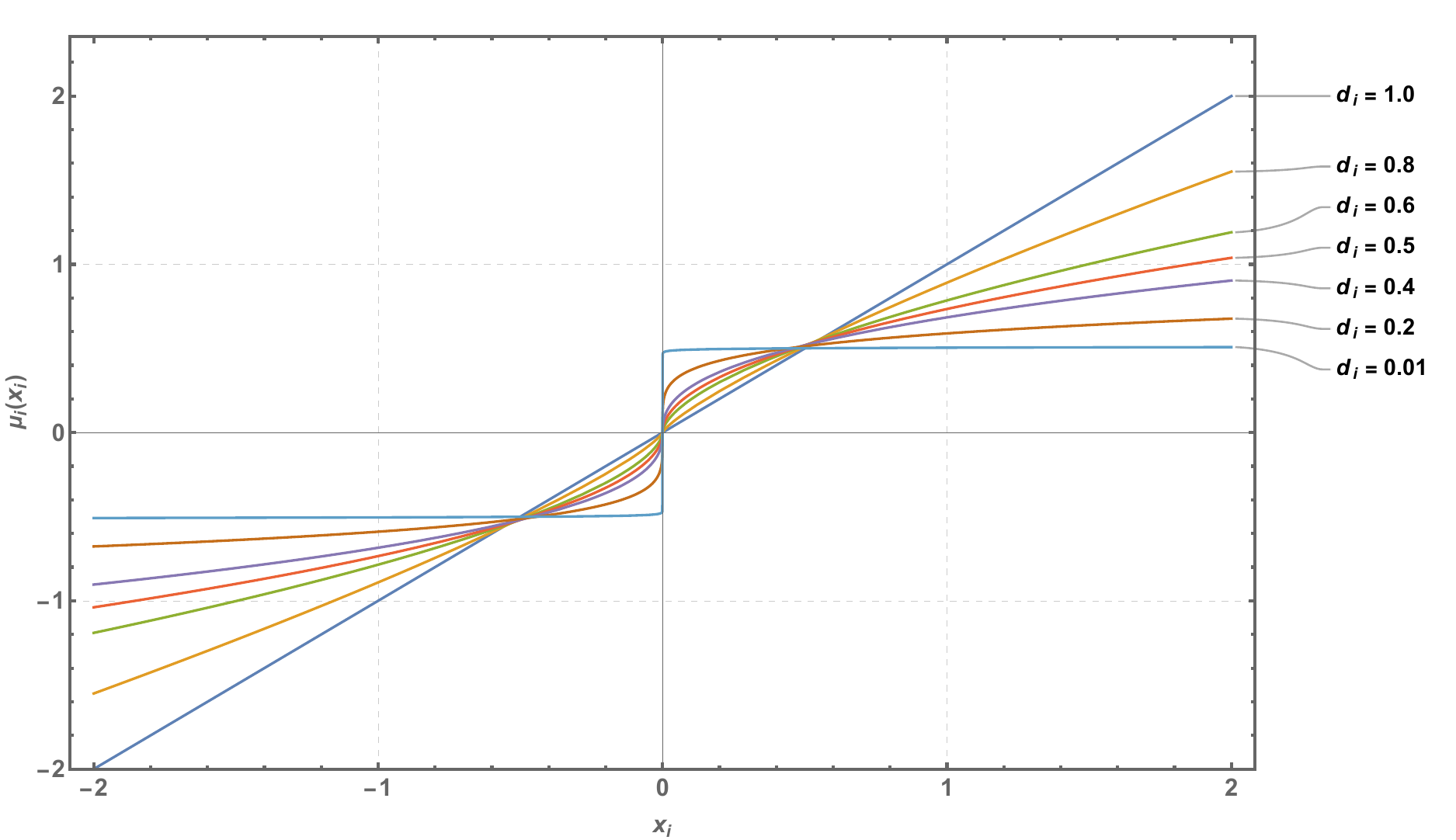}
	}
	\caption{FDG relation between the ``fractional'' coordinate $\mu _{i}$ and the ``standard'' coordinate $x_{i}$, for a fixed scale length $l_{i}=1$ and a variable fractional dimension $d_{i}$.}
	\label{figure:MWfig4}
\end{figure}

A simple integration of Eq. (\ref{eq5.1}) yields:\footnote{For simplicity, we will assume that the origins of the two systems coincide, i.e., we will not include any relative distance between their origins.} 
\begin{equation}\mu _{i}(x_{i}) =\frac{\pi ^{d_{i}/2}}{d_{i} \Gamma (d_{i}/2)}\left \vert \frac{x_{i}}{l_{i}}\right \vert ^{d_{i} -1} x_{i} ,\ i =1 ,2 ,3, \label{eq5.2}\end{equation}
where $\mu _{i}$ can be considered the ``fractional'' coordinates, as opposed to the ``standard'' $x_{i}$ coordinates. The former reduce to the latter in the case of $d_{i}=1$, but for $d_{i}<1$ there is a noticeable difference which is illustrated in Fig. \ref{figure:MWfig4}. In this figure, we plot the $\mu _{i}(x_{i})$ function for a fixed value of $l_{i}=1$ and different values of $d_{i}$. For $d_{i} = 1$, $\mu _{i}$ reduces to $x_{i}$, but for  $d_{i} < 1$ the differences in the two coordinates can be seen in the figure.

Typically, fractal behavior is considered for distances larger than the scale length, i.e. for $x_{i}\gtrsim l_{i}$, where the value of $l_{i}=1$, chosen for Fig. \ref{figure:MWfig4}, is shown by gray grid-lines in this figure. Assuming that both axes are using the same units (meters, kilo-parsecs, etc.), distances measured with the ``fractional'' coordinate $\mu _{i}$ appear to be shortened compared to the equivalent distances measured with $x _{i}$, for $d_{i}<1$ and $x_{i}\gtrsim l_{i}$.

This analysis will immediately impact the definition of displacement, velocity, acceleration, and all the other standard kinematic quantities. For example, from the ``standard'' velocity $v_{i}\left ( t \right )=\frac{dx_{i}}{dt}$ we can introduce the ``fractional'' velocity $\nu _{i}\left ( t \right )=\frac{d\mu_{i}}{dt}$, and using Eq. (\ref{eq5.1}) obtain:
\begin{equation}\nu _{i}\left ( t \right )=\frac{d\mu_{i}}{dt} =\frac{\pi ^{d_{i}/2}}{\Gamma (d_{i}/2)}\left \vert \frac{x_{i}(t)}{l_{i}}\right \vert ^{d_{i} -1}v_{i}(t) ,\ i =1 ,2 ,3. \label{eq5.3}\end{equation}

In this equation and in the following, we will assume that the time variable $t$ is not affected by any ``fractional'' effect, as opposed to the space variables. In our FDG model, we did not see any reason to consider the time variable as having any fractional/fractal behavior, so we will keep using standard time $t$ in this section.

In Newtonian kinematics, a simple motion with constant velocity $v_{i}$ in the $i$ direction is described by $x_{i}\left ( t \right )=x_{i}\left ( 0 \right )+v_{i}t$, where $x_{i}\left ( 0 \right )$ is the initial position at time $t=0$. We can assume that a similar motion with constant velocity, using ``fractional'' coordinates, is described by:
\begin{equation}\mu _{i}\left ( t \right )=\mu _{i}\left ( 0 \right )+\nu_{i}t, \label{eq5.4}\end{equation}
where $\nu_{i}$ is a constant ``fractional'' velocity, related to the standard velocity $v_{i}$ through Eq. (\ref{eq5.3}).

Using Eqs. (\ref{eq5.2})-(\ref{eq5.4}) and after some algebraic manipulation, it is possible to rewrite the constant velocity motion in Eq. (\ref{eq5.4}) as follows:
\begin{equation}x_{i}\left ( t \right )=\left (\pm l_{i} \right ) \left [\frac{d_{i}\Gamma \left ( d_{i}/2 \right ) }{\pi ^{d_{i}/2}\left (\pm l_{i}  \right )} \left (\mu _{i}\left ( 0 \right )+\nu _{i}t  \right ) \right ]^{1/d_{i}}, \label{eq5.5}\end{equation}
with $\mu _{i}(0) =\frac{\pi ^{d_{i}/2}}{d_{i} \Gamma (d_{i}/2)}\left \vert \frac{x_{i}(0)}{l_{i}}\right \vert ^{d_{i} -1} x_{i}(0)$, in view of Eq. (\ref{eq5.2}), and $\nu _{i}=\nu _{i}(0)=\frac{\pi ^{d_{i}/2}}{\Gamma (d_{i}/2)}\left \vert \frac{x_{i}(0)}{l_{i}}\right \vert ^{d_{i} -1}v_{i}(0)$, in view of Eq. (\ref{eq5.3}).\footnote{In Eq. (\ref{eq5.5}), $\left (+ l_{i} \right )$ will be used for $\left (\mu _{i}\left ( 0 \right )+\nu _{i}t  \right ) \ge 0$, and $\left (- l_{i} \right )$ will be used for $\left (\mu _{i}\left ( 0 \right )+\nu _{i}t  \right ) < 0$, respectively.}

\begin{figure}\centering 
 	\setlength\fboxrule{0in}\setlength\fboxsep{0.1in}\fcolorbox[HTML]{000000}{FFFFFF}{\includegraphics[width=6.50in, height=4.50in,]{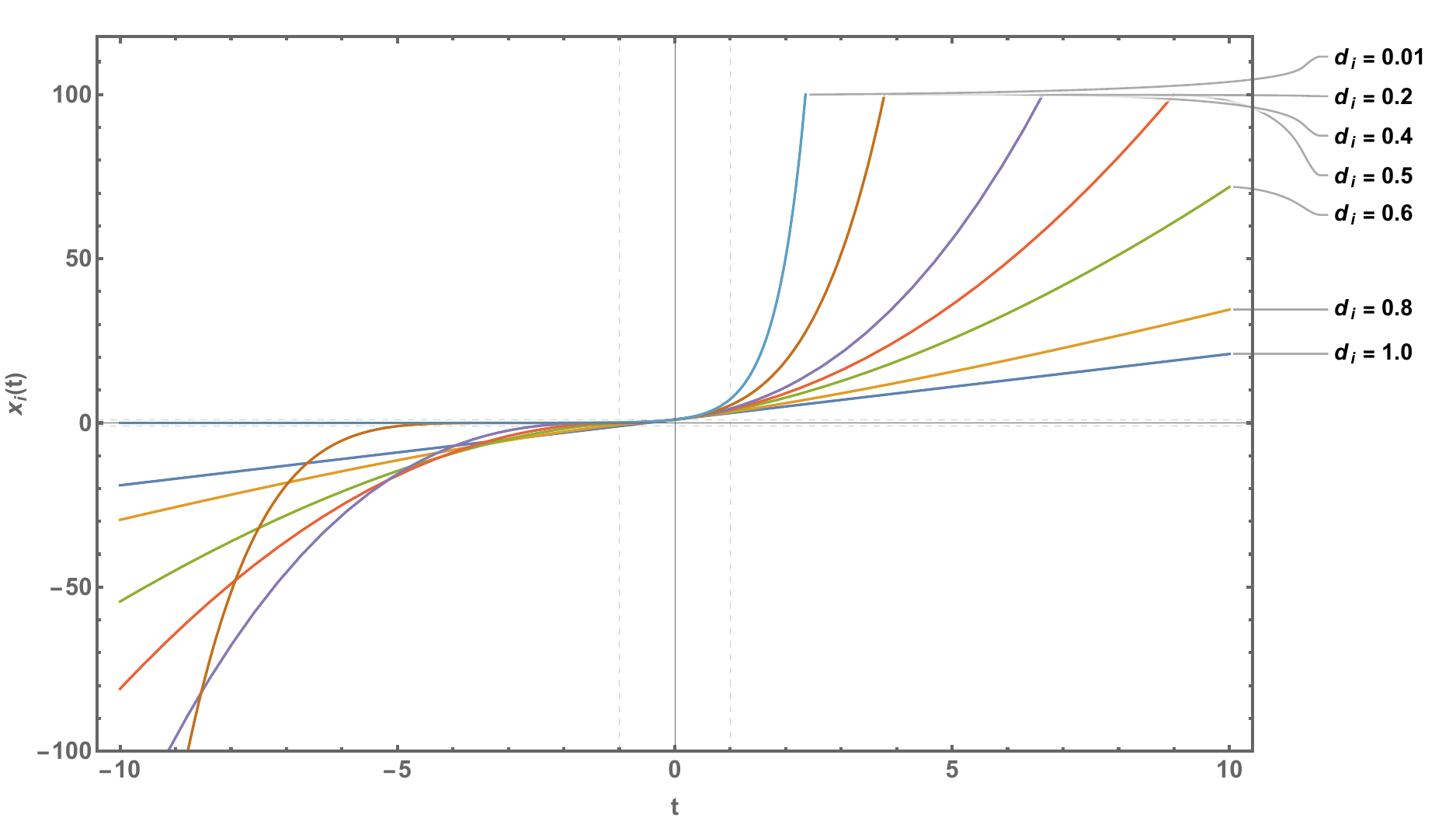}
 	}
 	\caption{FDG description in ``standard'' coordinates of a uniform velocity motion in ``fractional'' coordinates, for $l_{i}=1$, $x_{i}(0)=1$, $v_{i}=2$, and a variable fractional dimension $d_{i}$.}
 	\label{figure:MWfig5}
 \end{figure}

It is easy to check that Eq. (\ref{eq5.5}) reduces to standard $x_{i}\left ( t \right )=x_{i}\left ( 0 \right )+v_{i}t$ for $d_{i}=1$, but for $d_{i}<1$ a motion with constant fractional velocity $\nu_{i}$ will not correspond to a motion with constant standard velocity $v_{i}$. This is shown in Fig. \ref{figure:MWfig5}, where we set $l_{i}=1$, $x_{i}(0)=1$, $v_{i}=2$, and we show results for various values of $0<d_{i} \leq 1$ as was done in Fig. \refeq{figure:MWfig4}.

While for $d_{i}=1$ the plot of this motion with constant velocity $\nu_{i}$ is linear as expected, the plots for values $d_{i}<1$ are not linear and they show increasing slope for $\lvert x_{i} \rvert \gtrsim l_{i}$. This means that the ``standard'' velocity $v_{i}$ would be changing, thus these motions would not be uniform in standard coordinates.

Similar considerations would apply to the definition of a ``fractional'' acceleration $\alpha_{i}$. This acceleration can also be expressed in terms of standard quantities, by taking the time derivative of Eq. (\ref{eq5.3}):
\begin{equation}\alpha _{i}\left ( t \right ) = \frac{d\nu_{i}}{dt}  = \frac{d^{2}\mu_{i}}{dt^{2}} = \frac{\pi ^{d_{i}/2}}{\Gamma (d_{i}/2)}\left \vert \frac{x_{i}(t)}{l_{i}}\right \vert ^{d_{i} -1} \left [ \left ( d_{i}-1 \right ) \frac{v^{2}_{i}\left ( t \right )}{x_{i}\left ( t \right )} + a_{i}\left ( t \right ) \right ] ,\ i =1 ,2 ,3, \label{eq5.6}\end{equation}
which reduces to standard $a_{i}$ for $d_{i}=1$. 

Using $\alpha_{i}$, we could consider motions with constant ``fractional'' acceleration and other kinematic equations. Assuming the mass $m$ of a body to be a scalar, invariant quantity and introducing some equivalent of the Newtonian force $F$ in fractional coordinates, the laws of dynamics could also be extended to fractal/fractional spaces.

It is beyond the scope of this current work to pursue a fractional generalization of Newtonian dynamics, and we will leave this to future work on the subject. However, we will consider here a possible extension of Special Relativity, in view of the above discussion. We start by assuming that Einstein's two postulates of SR \cite{Griffiths_2017} (the principle of relativity, and the principle of invariant speed of light in vacuum) remain valid in any fractional spacetime, with space dimensions $\left\{ d_{i} \right\}, i=1,2,3$. 

The first principle would define inertial frames of reference as in standard dynamics, but considering only frames of references with the same fractal dimensions $\left\{ d_{i} \right\}, i=1,2,3$. This is due to the discussion following Eqs. (\ref{eq5.4})-(\ref{eq5.5}), regarding motions with constant velocity $\left\{ \nu_{i} \right\}$, which are not transformed into constant velocity motions, if the dimension of the space is changed.

We will assume that the second principle (constancy of the speed of light $c$ in inertial systems) remains valid also in fractional spacetimes, although all experimental tests of this principle (Michelson-Morley experiment and others) have been conducted in a 3-dimensional space. 

However, we will postulate that this principle remains valid also for ``fractional'' inertial systems, and extend the Lorentz transformations of SR between two fractional systems $S$ and $S'$ as follows:
\begin{gather}\mu_{x}' = \gamma (\mu_{x}-\nu t) \label{eq5.7} \\
\mu_{y}' =\mu_{y} \nonumber \\
\mu_{y}' =\mu_{y}\nonumber \\
t' = \gamma (t-\nu \mu_{x}/c^{2}). \nonumber \end{gather}

These transformations assume that the fractional space coordinates $\mu_{i}$ and $\mu_{i}'$ have the same dimensions $d_{i}~ (i=1,2,3)$, that the two systems coincide at $t=t'=0$, and the relative motion happens along the common $\mu_{x},\mu_{x}'$ direction. The gamma factor is here defined as $\gamma=\sqrt{1-\nu^{2}/c^{2}}$, where $\nu$ is the fractional relative velocity between the two fractional inertial systems $S$ and $S'$.

These assumptions would lead to a fractional extension of SR, and possibly also of GR, that will be left to future studies. As a final example of this possible extension of SR, we will consider the fractional analysis of a motion of a point particle of mass $m$ under a constant force in SR. Relativistic dynamics (see for example Sect. 12.2.4 in Ref. \cite{Griffiths_2017}) can be based on Newton's second law, $\mathbf{F}=\frac{d\mathbf{p}}{dt}$, but adopting the relativistic momentum $\mathbf{p}=\frac{m\mathbf{u}}{\sqrt{1-\frac{u^{2}}{c^{2}}}}$, where $\mathbf{u}$ denotes here the particle velocity.

For a constant force $F$ in a certain direction of the motion, solving for the velocity $u$ gives: $u=\frac{\left ( F/m \right )t}{\sqrt{1+\left ( Ft/mc \right )^{2}}}$, assuming that the particle starts from rest at the origin, at time $t=0$. An additional integration yields the standard position vs. time function of this motion in SR \cite{Griffiths_2017}:
\begin{equation}x\left ( t \right )=\frac{mc^{2}}{F}\left [ \sqrt{1+\left ( Ft/mc \right )^{2}} -1 \right ], \label{eq5.8}\end{equation}
which yields a ``hyperbolic'' motion, as opposed to a classical ``parabolic'' motion, when the space variable $x$ is plotted against time $t$. This shows graphically that in SR the particle velocity can only asymptotically approach the speed of light $c$, but will never exceed it.

However, in view of our fractional extension of SR in Eq. (\ref{eq5.7}) and related  discussion, we could re-compute the SR motion under a constant force and obtain the same solution in Eq. (\ref{eq5.8}), but for a fractional coordinate $\mu$:
\begin{equation}\mu\left ( t \right )=\frac{mc^{2}}{F}\left [ \sqrt{1+\left ( Ft/mc \right )^{2}} -1 \right ], \label{eq5.9}\end{equation}
where $F$ is some constant ``fractional'' force, equivalent to the standard constant force $F$.

Combining together Eq. (\ref{eq5.2}) and Eq. (\ref{eq5.9}), we obtain the position vs. time function in standard coordinates:
\begin{equation}x\left ( t \right )=l_{x}\left\{\frac{d_{x}\Gamma \left ( d_{x}/2 \right )}{\pi ^{d_{x}/2} l_{x}} \frac{mc^{2}}{F}\left [ \sqrt{1+\left ( Ft/mc \right )^{2}} -1 \right ]\right\}^{1/d_{x}}, \label{eq5.10}\end{equation}
and the related standard velocity $u(t)$ function can also be easily computed:
\begin{equation}u\left ( t \right )=\frac{\left ( \frac{F}{m} \right ) t}{\sqrt{1+\left ( Ft/mc \right )^{2}}}\frac{\Gamma\left ( d_{x}/2 \right ) }{\pi^{d_{x}/2} }\left\{\frac{d_{x}\Gamma \left ( d_{x}/2 \right )}{\pi ^{d_{x}/2} l_{x}} \frac{mc^{2}}{F}\left [ \sqrt{1+\left ( Ft/mc \right )^{2}} -1 \right ]\right\}^{\frac{1-d_{x}}{d_{x}}}. \label{eq5.11}\end{equation}

\begin{figure}\centering 
	\setlength\fboxrule{0in}\setlength\fboxsep{0.1in}\fcolorbox[HTML]{000000}{FFFFFF}{\includegraphics[width=6.50in, height=9.00in,]{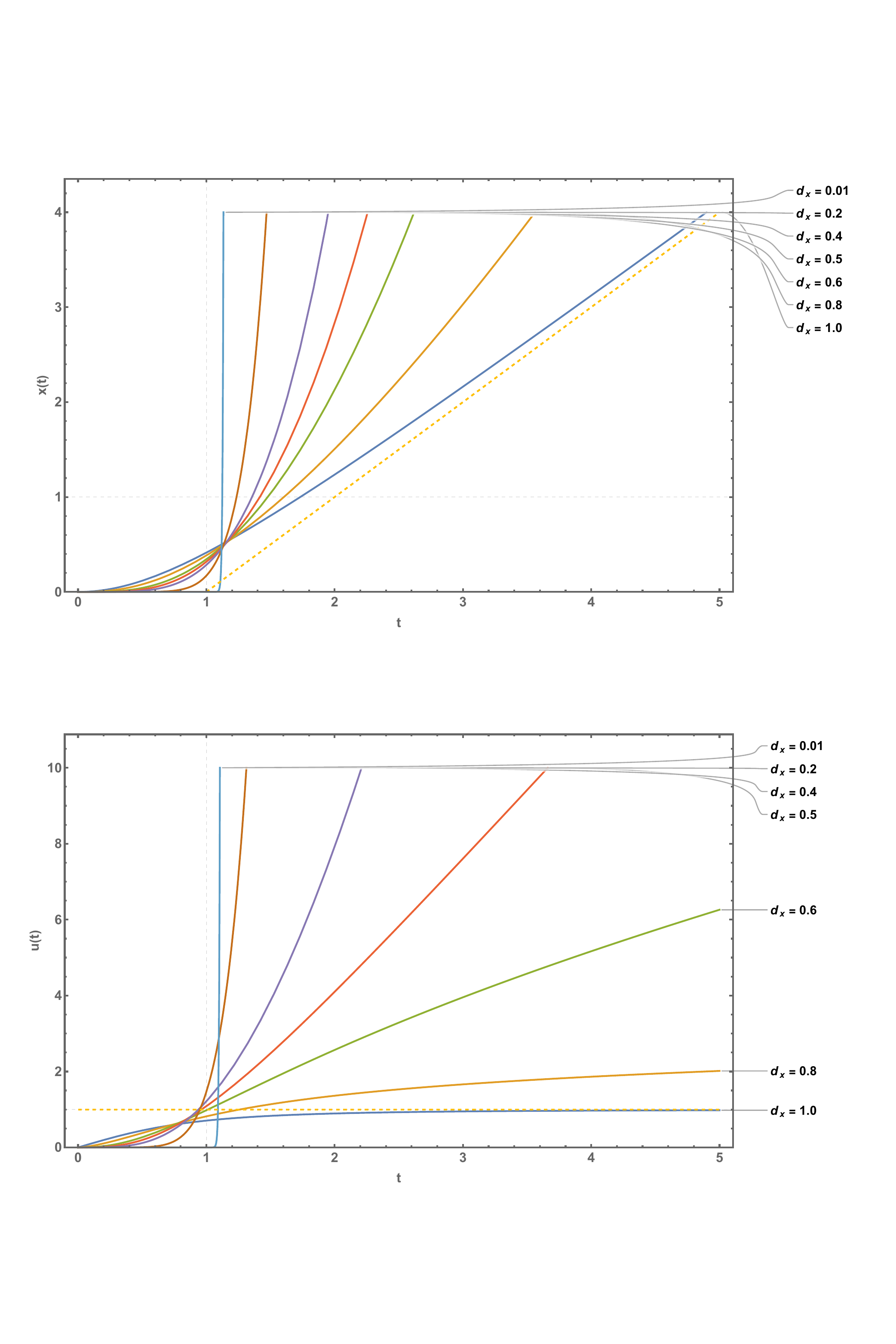}
	}
	\caption{FDG description in ``standard'' coordinates of a motion under a constant force $F$ in a ``fractional'' extension of SR. We assume that a point particle starts at rest at the origin, and consider its motion for fixed values of $F=m=c=l_{x}=1$ and variable values of the space dimension $d_{x}$. Top panel: position function $x(t)$, following Eq. (\ref{eq5.10}). Bottom panel: velocity function $u(t)$, following Eq. (\ref{eq5.11}).}
	\label{figure:MWfig6}
\end{figure}

These two functions $x(t)$ and $u(t)$ are plotted in Fig. \ref*{figure:MWfig6}, for fixed values of $F=m=c=l_{x}=1$ and variable values of the space dimension $d_{x}$. The curves for $d_{x}=1$ represent the standard SR results for a motion of a point particle under a constant force $F$: the $x(t)$ curve approaches asymptotically the worldline of a light signal (slope $=c=1$, dashed line in top panel), while the corresponding $u(t)$ curve shows that the speed of the particle is always less than $c$ ($c=1$, dashed line in bottom panel).

The curves for $d_{x}<1$ represent instead the same motion under a constant force $F$ in a fractional space, but as measured by an observer using standard space coordinates. In these cases, for large enough values of the time variable $t$, the object is seen moving with speed greater than $c$, if the observer uses standard coordinates. In other words, the object speed would still be limited by $c$ in the fractional inertial frame of reference, but it would become effectively superluminal as seen by an observer who assumes that $d_{x}=1$ and employs corresponding standard coordinates to measure the motion.

In Sect. \ref{sect:galactic}, we estimated the fractional dimension of the MW in the outer regions of the Solar System to be around $D_{SS} \approx 2.5-2.6$. Assuming that this reduced value of the space dimension is due to just one spatial direction and identifying it with the $x$ coordinate used in this section, we would have $d_{x} \approx 0.5-0.6$ for this particular direction, which could represent the distance from the center of the Solar System outwards.

Therefore, an object, such a spacecraft moving under a constant propulsion force, transitioning from the central $D=3$ region of the Solar System toward neighboring $D_{SS} \approx 2.5-2.6$ regions, would appear to be moving following the $d_{x} \approx 0.5-0.6$ curves in Fig. \ref{figure:MWfig6}, thus becoming effectively superluminal at some point during its motion. This effect is ultimately due to Eqs. (\ref{eq5.1})-(\ref{eq5.2}), which imply that distances measured with ``fractional'' coordinates are smaller than standard distances, for dimension $d_{i}<1$. Since the cosmic distance ladder used in astronomy and cosmology to estimate distances to celestial objects is based on standard $D=3$ space, FDG would also imply reconsidering most of the steps in this astronomical calibration.

We remark again that all the considerations in this section are very speculative and just follow from the mathematical theory of fractional-dimension spaces, without any experimental evidence apart from the fitting of galactic RCs done with FDG. Future work will improve and extend this preliminary analysis into a more comprehensive treatment of relativity and fractional-dimension gravity.  

\section{\label{sect:Conclusion}Conclusions}

In this work, we applied our FDG model and related computations to the Milky Way galaxy. We were able to show that our model is again effective in fitting the rotation curves for this galaxy without any DM contributions, provided we assume a fractional dimension, $D(R)$ or $D_{m}(R)$, as a function of the galactocentric radial distance.

Two different RC data were used: a unified RC compiled by Sofue over a large range of radial distances and a more recent set of data compiled by Sylos Labini over a shorter radial range, but using the latest Gaia DR3 data. Our general method employing the $D(R)$ function was highly effective for both data sets, yielding a fractional dimension $D \approx 2.3-2.6$ over most of the radial range.

The alternative method, based on the mass-dimension fractional dimension $D_{m}(R)$, was less effective in fitting the data of the extended RC, but still captured the general pattern of the curve. Both methods confirmed that the MW behaves like other previously studied galaxies, similarly characterized by a strong central bulge component and a dominating stellar disk at larger radial distances. We also noted that, at the radial position of the Solar System in the MW, the value of the fractional dimension can be estimated as $D_{SS} \approx 2.5-2.6$, meaning that the fractional dimension might transition from standard $D=3$ inside the Solar System to lower values $D_{SS} \approx 2.5-2.6$ just outside our solar neighborhood.  

The successful reproduction of the MW rotation curves with FDG methods, as well as those of previously studied galaxies \cite{Varieschi:webpage}, is certainly encouraging and worthy of further studies. However, the phenomenology usually attributed to DM extends substantially beyond galactic rotation curves, including structure formation on cosmological scales, gravitational lensing, galaxy and halo morphology, cluster dynamics, and CMB-scale effects. 

At present, it is not clear how FDG could evolve to address large scale structure formation and the three-dimensional morphology and dynamics of galaxies. One possible approach involves extending FDG from a heuristic, spherically or axially symmetric static framework into a dynamic, fully covariant, and field-theoretic model. A first attempt at this task was introduced in our relativistic FDG model \cite{Varieschi:2021rzk}, which may require further extension by adding a time-dependent weight factor, or a metric tensor that operates in fractional-dimension spacetimes. 

This would allow for the derivation of cosmological field equations to govern the global expansion rate as in standard cosmology. As for the complex morphology and dynamics of galaxies, the fractional dimension $D$ should be determined dynamically by a universal tensor or scalar field equation, rather than being hand-fitted to specific geometries with spherical or axial symmetry. These and other related topics will be addressed in future work on the subject.

As an initial step in this direction, in this paper we considered possible modifications to relativistic kinematics and dynamics in the framework of FDG. We performed a simplified analysis based on the one-dimensional fractional metric $d\mu_{i}$ and found interesting connections between fractional coordinates $\mu_{i}$ and standard coordinates $x_{i}$. This led to a simple generalization of other kinematic quantities, such as velocities and accelerations. 

Assuming also a simple fractional extension of the Lorentz transformations of Special Relativity, we briefly studied the case of a motion of a point mass under a constant force, in a one-dimensional space characterized by a fractional dimension $0<d_{x}<1$. The main consequence of this analysis is the possibility of an effective superluminal motion, as seen by an observer measuring the motion with standard coordinates. However, at the present stage, this is just a speculative possibility and by no means a demonstrated physical mechanism allowing material objects or information to propagate superluminally. 

Further studies will be needed to improve these extensions of FDG to relativistic kinematics and dynamics. In any case, the fractional-dimension gravity approach will continue to be relevant in astrophysics, gravitation, and cosmology, as an alternative theoretical framework to the dark matter paradigm. FDG might also be able to address other issues of standard $\Lambda CDM$ cosmology, such as the accelerated expansion of the universe and related dark energy problem.

\begin{acknowledgments}This study was supported by the Seaver College of Science and Engineering, Loyola Marymount University - Los Angeles. The author also wishes to thank the anonymous reviewer for useful comments and suggestions.
\end{acknowledgments}

\section*{Data Availability}

The data underlying this article will be shared on reasonable request to the corresponding author.

\bibliographystyle{apsrev4-2}
\bibliography{NFDG2025mainNotes}

\end{document}